\documentclass[11pt]{article}
\pdfoutput=1
\usepackage{jheppub}
\usepackage{slashed}
\usepackage{amsfonts,amsmath,amssymb,amsthm}
\usepackage{feynmp-auto}

\usepackage{latexsym}
\usepackage{graphicx}
\usepackage{slashed}
\usepackage[]{hyperref}
\hypersetup{colorlinks=true,linkcolor=magenta,anchorcolor=green,citecolor=cyan,filecolor=black,menucolor=black,urlcolor=black}
\numberwithin{equation}{section}

\def\stamp{--- {\bf \today} --- {\bf \jobname.tex}}

\def\sla#1{\not\!{#1}}

\def\tr{\textrm{tr}}

\def\cM{\mathcal{M}}
\def\cN{\mathcal{N}}

\def\Ree{\Re\textrm{e}}
\def\Imm{\Im\textrm{m}}

\newcount\hh
\newcount\mm
\mm=\time
\hh=\time
\divide\hh by 60
\divide\mm by 60
\multiply\mm by 60
\mm=-\mm
\advance\mm by \time
\def\thistime{\number\hh:\ifnum\mm<10{}0\fi\number\mm}
\def\stamp{--- {\bf \today} --- {\bf \thistime} --- {\bf \jobname.tex}}

\def\tr{\textrm{tr}}

\def\tr{\textrm{tr}}

\def\sign(#1){\textrm{sign}(#1)}

\def\cN{\mathcal{N}}

\def\cM{\mathcal{M}}

\def\ra{\rangle}

\def\BE{\begin{equation}}
\def\EE{\end{equation}}

\def\sla#1{\not\!{#1}}

 \def\<#1|#2){\left\langle#1|#2\right\rangle}
 \def\<#1|#2|#3]{\left\langle#1|#2|#3\right ]}
\def\(#1|#2|#3>{\left[#1|#2|#3\right\rangle}
 \def\[#1|#2]{\left[#1|#2\right]}

\def\an[#1,#2]{\left\langle#1\,#2\right\rangle}
\def\aq[#1,#2,#3]{\left\langle#1|#2|#3\right]}
\def\qa[#1,#2,#3]{\left[#1|#2|#3\right\rangle}
\def\sq[#1,#2]{\left[#1\,#2\right]}
\def\spa#1.#2{\left\langle#1\,#2\right\rangle}
\def\spab[#1,#2,#3]{\left\langle#1|#2|#3\right]}
\def\spba[#1,#2,#3]{\left[#1|#2|#3\right\rangle}
\def\spb#1.#2{\left[#1\,#2\right]}
\def\lor#1.#2{\left(#1\,#2\right)}

\begin{document}
\preprint{IPHT-t16/082, ACFI-T16-23}
\title{\LARGE Light-like Scattering in Quantum Gravity\smallskip}\
\author[a]{N.~E.~J.~Bjerrum-Bohr}
\author[b]{ John~F.~Donoghue}
\author[b,c]{Barry~R.~Holstein}
\author[d]{Ludovic~Plant\'e} 
\author[d]{Pierre~Vanhove}
\affiliation[a]{\small Niels Bohr  International  Academy \&  Discovery
Center, Niels\,Bohr\,Institute,\,Blegdamsvej\,17,\,DK-2100,\,Copenhagen\,\O,\,DENMARK\smallskip}
\affiliation[b]{Department of Physics-LGRT, University of Massachusetts,
Amherst, MA 01003, USA}
\affiliation[c]{Kavli Institute for Theoretical Physics, University of California, Santa Barbara, CA  93016}
\affiliation[d]{CEA,\,DSM,\,Institut\,de\,Physique\,Th{\'e}orique,\,IPhT,\,CNRS\,MPPU,
\,URA2306,\,Saclay,\,F-91191,\,Gif-sur-Yvette,\,FRANCE
}

\date{\today}\bigskip

\abstract{
%\vspace{0.7cm}
We consider scattering in quantum gravity and derive long-range classical and quantum
contributions to the scattering of light-like bosons and fermions
(spin-0, spin-$\frac12$, spin-1) from an external massive scalar
field, such as the Sun or a black hole.  This is achieved by treating
general relativity as an effective field theory and identifying the
non-analytic pieces of the one-loop gravitational scattering
amplitude.  It is emphasized throughout the paper how modern amplitude
techniques, involving spinor-helicity variables, unitarity, and
squaring relations in gravity enable much simplified computations.  We
directly verify, as predicted by general relativity, that all
classical effects in our computation are universal (in the context of
matter type and statistics). Using an eikonal procedure we confirm
the post-Newtonian general relativity correction for light-like bending around
large stellar objects. We also comment on treating effects from
quantum $\hbar$ dependent terms using the same eikonal method.
}
%\pacs{}

\maketitle
 \tableofcontents
%\newpage
%%%%%%%%%%%%%%%%%%%%%%%%%%%%%%%%%%%%%%%%%%%%%%%%%%%%%%%%%%%%%%%%%
%--------------------------------------------------------------------------

 \section{Introduction}
 \label{sec:introduction}
The possible existence of a quantum field theoretical framework for general relativity valid at
all energy scales is clearly a fundamental question, and since the original
formulation of quantum field theory, a technique by which general relativity and quantum mechanics can be combined has been sought~\cite{Fierz:1939ix,Gupta:1954zz,Kraichnan:1955zz,Feynman:1963ax,DeWitt:1967yk,DeWitt:1967ub,DeWitt:1967uc}.
Though such a theory has yet to be found, today we can address profound practical and reliable (low energy) consequences of the (currently unknown) underlying quantum theory through the modern viewpoint of effective field
theory (EFT)~\cite{Weinberg:1978kz,Donoghue:1993eb,Donoghue:1994dn,BjerrumBohr:2002sx,BjerrumBohr:2002ks,BjerrumBohr:2002kt,Khriplovich:2004cx,Holstein:2006pq,Ross:2007zza,Holstein:2008sx,Holstein:2008sy,Neill:2013wsa,Bjerrum-Bohr:2013bxa,Bjerrum-Bohr:2014lea,Bjerrum-Bohr:2014zsa,Bjerrum-Bohr:2015vda}. The EFT framework allows direct exploration of various quantitative phenomenological applications, see for example~\cite{Burgess:2003jk,Donoghue:2015hwa}.  In addition, the recent detection of gravitational waves GW150914 by the LIGO experiment is an important test of
general relativity~\cite{TheLIGOScientific:2016src} and opens up
exciting prospects for testing low-energy effective theories of
gravity~\cite{Yunes:2016jcc}.  In the analysis below we will be
seeking the  classical and quantum  long-range (power law falloff) corrections to the familiar
$1/r$ Newtonian potential describing the gravitational interaction
between two systems.  The effective potential describing this
interaction is defined as the Fourier transform of the gravitational
scattering amplitude and $1/r^n,\,n\geq 2$ behavior can only arise
from non-analytic components associated with quantized graviton loop
effects.  Analytic pieces lead only to short-distance (delta function
and its derivatives) behavior and can be dropped when only concerned
with long range physics.

In this paper  we focus on providing further details on the effective field theory computation of light-like scattering in quantum gravity.
In particular we will extend our previous results for light-like scattering from bosons to fermions. Thus we can now address massless gravitational neutrino scattering in one-loop quantum gravity.
The outline of our presentation will be as follows: We first discuss the framework for our computation; especially we will show how modern computational techniques, unitarity and spinor-helicity are important inputs for streamlining the computations. Next we will present details of the calculation, and finally discuss how to interpret our results.

%%%%%%%%%%%%%%%%%%%%%%%%%%%%%%%%%%%%%%%%%%%%%%%%%%%%%%%%%%%%%%%%
\section{General relativity as an effective field theory and one-loop amplitudes}

Including gravitational interactions in particle physics models is a straightforward exercise employing ideas from effective field theory. The starting point is the gravitational effective field theory action
\begin{eqnarray}\label{e:starting-point}
  \mathcal S&=& \int d^4x \,
  \sqrt{-g}\,\bigg [{2\over \kappa^2} \,\mathcal R  + S_{\text {model}}+ S_{\text {EF}}\bigg ]\,,
\end{eqnarray}
where $\mathcal R$ is the scalar curvature and $g_{\mu\nu}$ the metric. One can write the metric as
$\eta_{\mu \nu}+\kappa h_{\mu\nu}$, with $\kappa^2 = 32 \pi  G_N/c^4$, where $G_N$ is Newton's constant and
$h_{\mu\nu}$ is the quantized gravitational field.
Expanding all terms in $h_{\mu\nu}$, $\mathcal R$ contains the propagator for the gravitational field as well as all the pure gravitational vertices.
Interaction with matter is contained in the term
$S_{\text {model}}$ where the flat space Lagrangian for a given particle physics model is made generally covariant by replacing flat space derivatives with general covariant derivatives expanded in powers of $\kappa h_{\mu\nu}$ (see {\it e.g.} refs.~\cite{Donoghue:1995cz,BjerrumBohr:2004mz} for details regarding such expansions.).  Finally $S_{\text {EF}}$ contains
an infinite series of higher derivative operators (basically any
operator allowed by general covariance) associated with new
gravitational couplings, and ensures that, order by order in the
energy expansion, {\it any} UV divergence due to loop
effects~\cite{'tHooft:1974bx} can be absorbed in the effective
action. In this way the construction, albeit ``effective", is UV
consistent up to the cut-off determined by the validity of the energy
expansion, typically ${\cal O}(m_{\rm Planck})$, where $m_{\rm Planck}=\sqrt{\hbar c\over G}\sim 10^{19}$ GeV is the Planck mass.

Having constructed the effective action, we have now, in principle, a
straightforward path by which to derive transition amplitudes.  The
action term corresponding to the matter coupling $S_{\text{model}}$
can take different forms depending on the specific theory we wish to
study.  We will, in this presentation, need only the minimal couplings of gravitons to scalars, photons and massless fermions.

The action for a massless scalar $\varphi$ or massive scalar field $\Phi$  of mass $M$ is
\begin{equation}
  \mathcal S_{\text{scalar}} = \int d^4x \,
  \sqrt{-g}\,\left(-\frac12 (\partial_\mu \varphi)^2-\frac12
  \bigg((\partial_\mu\Phi)^2-M^2 \Phi^2\bigg)\right)\,,\label{e:EHscalar}
\end{equation}
while the coupling to a massless spin-${1\over 2}$ fermion is given by
\begin{equation}
  \mathcal S_{\text{fermion}} ={i\over2} \int d^4x \,
  \sqrt{-g}\,\bar\chi \slashed{D} \chi\,,\label{e:EJscalar}
\end{equation}
where $\slashed{D}= \gamma^\mu \,(\partial_\mu + {i\over4}
\omega_\mu^{ab} \gamma^{ab})$, $\omega_\mu^{ab}$ is the spin
connection, and $\gamma^\mu$ are 4$\times$4 Dirac matrices with
$\gamma^{ab}=\frac12[\gamma^a,\gamma^b]$ the Lorentz generator.
Finally, the coupling to an Abelian spin-1 massless field is given by
\begin{equation}
  \mathcal S_{\text{QED}} =-\frac14 \int d^4x \,
  \sqrt{-g}\,\left(\nabla_\mu
    A_\nu-\nabla_\nu A_\mu\right)^2\,,\label{e:QED}
\end{equation}
where $\nabla_\mu A^\nu:=\partial_\mu
    A^\nu+\Gamma^\nu{}_{\mu\lambda} A^\lambda$ and  $\Gamma^\lambda{}_{\mu\nu}:=\frac12 \,
g^{\lambda\sigma}(\partial_\mu g_{\sigma\nu}+\partial_\nu
g_{\sigma\mu}-\partial_\sigma g_{\mu\nu}) $. A full list of propagators and
vertices needed for the Feynman graphs computation can, {\it e.g.}, be found
in~\cite{Donoghue:1995cz,BjerrumBohr:2004mz,BlackburnThesis,Holstein:2008sx}.

Having an effective action, the traditional way to proceed is to simply work out all necessary Feynman rules to a particular loop order and then generate amplitudes perturbatively using the standard off-shell formalism. This, for instance, was the path taken in
refs.~\cite{BjerrumBohr:2002sx,BjerrumBohr:2002ks,BjerrumBohr:2002kt}.
However, at high orders such an approach is clearly not very practical, since it quickly leads to very unmanageable computations, in part because of vast off-shell vertices and the occurrence of tensor contractions everywhere. It is thus natural to take advantage of {\it any} available simplification and the on-shell techniques provided by the compact formalism of spinor-helicity and unitarity seems particularly ideal in this regard~\cite{Mangano:1990by,Dixon:1996wi,Dixon:2013uaa}. Also, importantly, recent progress in the computation of gluon and QCD amplitudes (see {\it e.g. }~\cite{Bern:1994cg,Witten:2003nn,Britto:2005fq,Berger:2008sj}) can be
adapted to gravity~\cite{Dunbar:1995ed,Bern:2002kj} using the Kawai-Lewellen-Tye (KLT) string theory relations~\cite{Kawai:1985xq,BjerrumBohr:2010hn}.
Using these methods, the only required input for effective field theory computations is that of compact on-shell tree amplitudes, since loop amplitudes can be written in terms of trees by the use of unitarity as the central consistency requirement.

To illustrate how this program is carried out in practice, we follow the procedure outlined in ref.~\cite{Bjerrum-Bohr:2014lea} where we considered two-graviton-exchange between massive scalar sources. The only difference between that approach and the present one is that we here extend the analysis to two-graviton-exchange between a massless
field $X$ with spin $S$ and a massive scalar field $\Phi$ with spin
$S=0$ and mass $M$.  Given the necessary tree amplitudes, we compute
the discontinuity across the two-particle $t$-channel cut via
\begin{eqnarray}
  \label{e:oneloop}
 \cM^{(2)}_{X}(p_1,p_2,p_3,p_4) \Big|_{\rm disc}:={1\over 2!\, i}
  \mu^{2\epsilon}\int d{\rm LIPS}(\ell_1,-\ell_2)\, (2\pi)^4\delta^4(p_1+p_2+p_3+p_4)\,\nonumber\\
  \times\sum_{\lambda_1,\lambda_2}  \cM^{(1)}_{X^2 G^2}(p_1,\ell_1,p_2-\ell_2) \,\times\,
  \cM^{(1)}_{\phi^2G^2}(p_3,\ell_2,p_4,-\ell_1)^\dagger\,.
\end{eqnarray}
where $d{\rm LIPS}(\ell_1,-\ell_2)= d^4\ell_1\,d^4\ell_2
\delta^{(+)}(\ell_1^2)
\delta^{(+)}(\ell_2^2)\,\delta(p_1+p_2+\ell_1-\ell_2)$. The amplitude
is related to the scattering matrix $T$ by 
$\langle p_1,p_2| i T|p_3,p_4\rangle=i \cM^{(2)}_X(p_1,p_2,p_3,p_4) \,
(2\pi)^4\, \delta^{(4)}(\sum_{i=1}^4 p_i)$.
We follow here the notation of ref.~\cite{Bjerrum-Bohr:2013bxa} and will
everywhere employ $D=4-2\epsilon$, defining $q:= p_1+p_2=-p_3-p_4=
\ell_2-\ell_1$, and $p_3^2:= p_4^2 := M^2$. We use
the mostly minus metric convention $( +, -, -, - )$. The Mandelstam variables
are $t:=q^2$, $s=(p_1+p_4)^2$ and $u=(p_1+p_3)^2$.

\begin{figure}[h]
  \centering
\begin{fmffile}{CutGraph}
    \begin{fmfgraph*}(200,100)
      \fmfleftn{i}{2}
      \fmfrightn{o}{2}
      \fmftop{v3}
      \fmfbottom{v4}
	\fmfrpolyn{smooth,filled=30,label=tree}{G}{4}
	\fmfpolyn{smooth,filled=30,label=tree}{K}{4}
\fmf{dashes_arrow,tension=2.5,label=$p_1$}{i1,G1}
\fmf{dashes_arrow,tension=2.5,label=$p_2$}{i2,G2}
\fmf{fermion,tension=2.5,label=$p_4$}{o1,K1}
\fmf{fermion,tension=2.5,label=$p_3$}{o2,K2}
\fmf{dbl_curly,left=.5,tension=.5,label=${\ell_2\atop \rightarrow}\hspace{2cm}$,label.side=left,label.dist=3}{G3,K3}
\fmf{dbl_curly,right=.5,tension=.5,label=$\hspace{2cm}{\leftarrow\atop\ell_1}$,label.side=right,label.dist=5}{G4,K4}
\fmf{dashes,fore=red}{v3,v4}
\end{fmfgraph*}
\end{fmffile}
  \caption{The one-loop scattering of  one
    massless scalars (dotted line) and one massive scalars (solid line) mediated
  by a graviton (curly line). The discontinuity cut is represented by
  the red line.}
  \label{fig:ggssoneloop}
\end{figure}
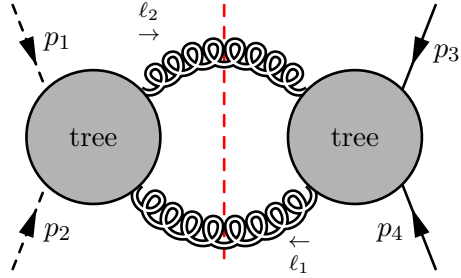

Here $\cM^{(1)}_{X^2 G^2}$ is the gravitational Compton amplitude
analyzed in~\cite{Bjerrum-Bohr:2014lea} (See also appendix~\ref{sec:gravtree}) and
the summation is over all possible helicity configurations $\lambda_1$ and
$\lambda_2$ across the cut.
%--------------------------------------------------------------------------
The two-particle cut can be pictorially represented as
in fig.~\ref{fig:ggssoneloop}
%
%--------------------------------------------------------------------------

On the
cut lines the gravitons are on-shell so that we have the constraint $\ell_1^2=0,\ \ell_2^2=0$.
The discontinuity is given by the sum of four box integrals with the same
numerator factor  (see
appendix~\ref{sec:gravtree}) 
\begin{equation}  \label{e:oneloops}
   \cM^{(2)}_{\varphi}(p_1,p_2,p_3,p_4)
  =-{\kappa^4\over 32 t^2\, i}\sum_{i=1}^2\sum_{j=3}^4
  \int {d^D\ell \,\mu^{2\epsilon}\over(2\pi)^D}\,{\mathcal N^{S}\over \ell_1^2\ell_2^2 (p_i\cdot\ell_1)(p_j\cdot\ell_1)}\,.
\end{equation}
With this construction one captures all the $t$-channel massless thresholds,
which are the only terms (the non-analytic ones) of interest to us.
Notice that the structure of the cut is very similar to that
evaluated in~\cite{Bjerrum-Bohr:2013bxa}.
The numerator $\mathcal N^{S}$ receives contributions from the singlet
graviton cut ({\it i.e.} where the helicities of the two cut gravitons
are identical) as well as from the non-singlet graviton cut ({\it i.e.} where the
helicities of the two cut gravitons are opposite).

\noindent $\bullet$ In the case of the massless external field, the properties of the
gravitational Compton amplitudes (reviewed in
appendix~\ref{sec:compton}) imply that
\begin{equation}
  \mathcal N^{S}_{\rm singlet}= 0\,,\qquad \textrm{for}~S=0,\,{\tfrac12},\,1\,.
\end{equation}
\noindent $\bullet$ The non-singlet cut for the massless scalar $S=0$ can be obtained by
applying the equations~(III.24) and~(III.27)
of~\cite{Bjerrum-Bohr:2013bxa}  with $m_1=0$ and $m_2=M$, yielding 
(we refer to the appendix~\ref{sec:conventions} for conventions and notations)\footnote{The computations performed in this work only involve parity even
    contributions, therefore the four-dimension  Levi-Civita epsilon tensor 
    $\varepsilon^{\mu\nu\rho\sigma}$  will never appear and there
    will be no issue in  evaluating of the one-loop amplitude using dimensional regularisation.
}
\begin{equation}\label{e:Nscalar}
  \cN^0_{\rm non-singlet}=\frac12\,\left[\big(\tr_-(\ell_1p_1\ell_2p_3)\big)^4+  \big(\tr_-(\ell_2p_1\ell_1p_3)\big)^4\right]\,.
\end{equation}

\noindent $\bullet$ For the photon, we denote the non-singlet cut by $ \mathcal
N_{\rm non-singlet}^{1\,h_1\, h_2}$, where
the polarization of the incoming photon  is $h_1$ and the polarization of
the outgoing photon is $-h_2$.
The only non-vanishing amplitudes are those preserving photon helicity
\begin{equation}
  \label{e:Nphoton}
  \mathcal  N_{\rm non-singlet}^{1\,+-}={\big(\tr_-(\ell_2 p_1\ell_1 p_3)
    \tr_+(\ell_2 p_3\ell_1 p_1p_3 p_2)\big)^2+(\ell_1\leftrightarrow \ell_2)\over \spab[p_1,p_3,p_2]^2}\,,
\end{equation}
and  $(\mathcal N^{1\,-+}_{\rm non-singlet})^*=\mathcal N^{1\,+-}_{\rm non-singlet}$.
Remarking that for the four-point amplitude the
$\epsilon^{\mu\nu\rho\sigma}$ terms do not contribute, we conclude that
\begin{equation}\label{e:photonNumRel}
\spab[p_1,p_3,p_2]^2 {\mathcal
  N}^{1\,+-}=\spab[p_2,p_3,p_1]^2\, {\mathcal N}^{1\,-+}=2\,\Ree\Big[\left(\tr_-(\ell_2 p_1\ell_1 p_3)
    \tr_+(\ell_2 p_3\ell_1 p_1p_3 p_2)\right)^2\Big]\,.
\end{equation}

\noindent $\bullet$ For the  massless fermion the non-singlet cut is non-vanishing as well
only for the helicity conserving case, and we have
\begin{equation}
  \label{e:Nfermion}
  \mathcal  N_{\rm non-singlet}^{\frac12\, +-}={\big(\tr_-(\ell_1 p_1\ell_2 p_3)^3
    \tr_+(p_1 p_3 p_2 \ell_1 p_3\ell_2 )\big)-(\ell_1\leftrightarrow \ell_2)\over \spab[p_2,p_3,p_1]}\,,
\end{equation}
and  $(\mathcal N^{\frac12\,-+}_{\rm non-singlet})^*=-\mathcal
N^{\frac12\, +-}_{\rm non-singlet}$. The polarization of the external
state appears only in the numerator, which takes the form
\begin{equation}\label{e:fermionNumRel}
    \spab[p_2,p_3,p_1]  \mathcal  N_{\rm non-singlet}^{\frac12\, +-}=
    \spab[p_1,p_3,p_2]  \mathcal  N_{\rm non-singlet}^{\frac12\,
      -+}=2i\, \Imm\Big[ (\tr_-(\ell_1 p_1\ell_2 p_3)^3
    \tr_+(p_1 p_3 p_2 \ell_1 p_3\ell_2 )\Big]\,,
\end{equation}
and, by multiplying by an appropriate factor, one can remove
this polarization dependence.  Therefore, for the
photon, we can define the coefficients as coming from the expansion of
\begin{equation}\label{e:photonAmplitude}
\cM_\gamma^{(2)}:=    \spab[p_1,p_3,p_2]^2 \mathcal
M_\gamma^{+-\,(2)}=\spab[p_2,p_3,p_1]^2\,\cM_\gamma^{-+\, (2)}\,,
\end{equation}
and similarly, for the fermion amplitude we define the coefficients from
the polarization-stripped expression
\begin{equation}\label{e:fermionAmplitude}
\cM_\chi^{(2)}:=    \spab[p_1,p_3,p_2] \mathcal
M_\chi^{+-\,(2)}=\spab[p_2,p_3,p_1]\,\cM_\chi^{-+\, (2)}\,.
\end{equation}

Performing the tensor integral reductions, the amplitude can be decomposed in terms of
integral functions containing the two-massless-particle $t$-channel cut
\begin{multline}\label{e:oneloopsInt}
\kappa^{-4}\,  \cM^{(2)}_{X}(p_1,p_2,p_3,p_4)\Big|_{t-cut}
= bo^S(t,s)   \, I_4(t,s)+
  bo^S(t,u)\, I_4(t,u)\cr
+ t_{12}^S(t) \, I_3(t,0)+ t_{34}^S(t)\, I_3(t,M^2)
+ bu^S(t,0) \, I_2(t,0)
\,,
\end{multline}
where $I_4(t,s)$ and $I_4(t,u)$ are scalar box integrals given in Eq.~(\ref{e:scalarbox}), $I_3(t)$ is
the massless triangle integral of Eq.~(\ref{e:masslesstriangle}),
$I_3(t,M^2)$ the massive triangle integral of
Eq.~(\ref{e:massivetriangle}), and $I_2(t)$
is the massless scalar bubble integral given in
Eq.~(\ref{e:masslessbubble}). Explicit expressions for these integrals
can be found in appendix~\ref{sec:integral}.
The {\it full} integral reduction gives, in addition, massive bubbles,
tadpoles, and as well as rational pieces that are (restricting to four
dimensions) {\it not} contained in the massless $t$-cut.  These terms are analytic in $t$ and are not of interest to our analysis.

\begin{figure}[ht]
  \centering
\begin{fmffile}{BCJGraph}
  \begin{equation*}
\begin{gathered}
    \begin{fmfgraph*}(150,100)
      \fmfleftn{i}{2}
      \fmfrightn{o}{2}
      \fmftop{v3}
      \fmfbottom{v4}
	\fmfrpolyn{smooth,filled=30,label=tree}{G}{4}
\fmf{plain}{o1,v1}
\fmf{plain}{o2,v2}
\fmf{plain,tension=.8}{v1,v2}
\fmf{dashes,tension=2.5}{i1,G1}
\fmf{dashes,tension=2.5}{i2,G2}
\fmf{dbl_curly,tension=.6}{G3,v2}
\fmf{dbl_curly,tension=.6}{G4,v1}
\fmf{dashes,fore=red,label={\vspace{-5cm}$(a)$}}{v3,v4}
\end{fmfgraph*}
\end{gathered}
\begin{gathered}
    \begin{fmfgraph*}(150,100)
      \fmfleftn{i}{2}
      \fmfrightn{o}{2}
      \fmftop{v3}
      \fmfbottom{v4}
	\fmfrpolyn{smooth,filled=30,label=tree}{G}{4}
\fmf{plain}{o1,v1}
\fmf{plain}{o2,v2}
\fmf{plain,tension=.1}{v1,v2}
\fmf{dashes,tension=2.5}{i1,G1}
\fmf{dashes,tension=2.5}{i2,G2}
\fmf{dbl_curly}{G3,v1}
\fmf{dbl_curly}{G4,v2}
\fmf{dashes,fore=red,label={\vspace{-5cm}$(b)$}}{v3,v4}
\end{fmfgraph*}
\end{gathered}
\begin{gathered}
    \begin{fmfgraph*}(150,100)
      \fmfleftn{i}{2}
      \fmfrightn{o}{2}
      \fmftop{v3}
      \fmfbottom{v4}
	\fmfrpolyn{smooth,filled=30,label=tree}{G}{4}
\fmf{plain}{o1,v1,o2}
\fmf{dashes,tension=2.5}{i1,G1}
\fmf{dashes,tension=2.5}{i2,G2}
\fmf{dbl_curly}{G4,v2,G3}
\fmf{dbl_curly,tension=3.5}{v2,v1}
\fmf{dashes,fore=red,label={\vspace{-5cm}$(c)$}}{v3,v4}
\end{fmfgraph*}
\end{gathered}
\end{equation*}
\end{fmffile}
  \caption{The  monodromy BCJ relations in the cut (in red) that link
    the one-loop integral coefficients.}
  \label{fig:bcj}
\end{figure}
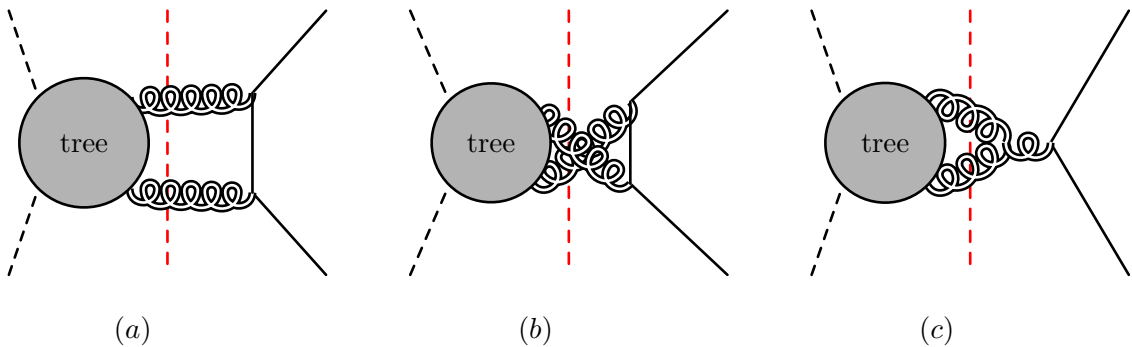

The massless triangle coefficient $t_{12}(t)$ is related to the
coefficients of the box $bo^S(t,s)$ and crossed-box $bo^S(t,u)$ by
\begin{equation}\label{e:BCJscalar}
  {bo^S(t,s)\over M^2-s}+{bo^S(t,u)\over M^2-u}= t^S_{12}(t)\,, \qquad \textrm{for}~S=0,\,\tfrac12,\,1\,.
\end{equation}
and this universal identity, which is a consequence of the  monodromy BCJ
relations~\cite{Bern:2008qj} between the four point tree-level amplitude
in the two-particle cut as depicted
in figure~\ref{fig:bcj}, is a very useful check on computations. The consequence of the monodromy BCJ
relations for one-loop integral coefficients have been studied in~\cite{Chester:2016ojq}
and~\cite{Primo:2016omk}, while a string theory based systematic derivation of these relations
was given in~\cite{Tourkine:2016bak}.

\section{The one-loop integral coefficients}

We now provide explicit expressions for the integral
coefficients of the one-loop amplitudes in Eq.~(\ref{e:oneloopsInt}) for
the massless scalar $X=\varphi$, the photon $X=\gamma$, and the
massless fermion $X=\chi$.

\medskip
\noindent {\it The box coefficients are given by }
\begin{itemize}
\item[ $\bullet$] for the scalar \smallskip \smallskip
\begin{equation}
  \label{e:scalarboxst}
\hskip-10cm bo^\varphi(t,s)=\frac14 \left(M^2-s\right)^4\,,
\end{equation}
and with $u$ replacing $s$ for the coefficient $bo^\varphi(t,u)$ of
the cross box.
\item[$\bullet$] for the photon
\begin{equation}\begin{split}
  \label{e:photonboxst}
\hskip-0.75cm bo^\gamma(t,s)&=  {\big(M^2-s\big)^2\over8}\, \big(2 M^8-8 M^6 s+2 M^4 s (t+6 s)-4 M^2 s^2 (t+2
   s)\\&+s^2 \big(t^2+2 t s+2 s^2\big)\big)\,,
\end{split}\end{equation}
and with $u$ replacing $s$ for the coefficient $bo^\gamma(t,u)$ of the
cross box.
\item[ $\bullet$]  for the fermion\smallskip\smallskip
\begin{equation}
  \label{e:fermionboxst}
\hskip-5.55cm bo^\chi(t,s)=  {\left(M^2-s\right)^3\over8}\, \left(2 M^4-4 M^2 s+s(t+2s)\right)\,,
\end{equation}
and with $u$ replacing $s$ for the coefficient $bo^\chi(t,u)$ of the
cross box.
\end{itemize}
%
%\break
\noindent {\it The massless triangle coefficients are given by }
\begin{itemize}
\item[ $\bullet$] for the scalar\smallskip\smallskip
\begin{equation}
  \label{e:scalartri12}
\hskip-7.85cmt_{12}^\varphi(t,s)= \frac14 (M^2-s)^3+\frac14 (M^2-u)^3\,,
\end{equation}
\item[$\bullet$] for the photon
\begin{multline}
  \label{e:photontri12}
t_{12}^\gamma(t,s)=  \frac{t}{8}\, \Big(6 M^8-2 M^6 (5 t+12 s)+2 M^4
\left(5 t^2+16 t s+18 s^2\right) \cr-M^2 \left(5
   t^3+20 t^2 s+34 t s^2+24 s^3\right)+\left(t^2+2 t s+2 s^2\right) \left(t^2+3 t s+3 s^2\right)\Big)\,,
\end{multline}
\item[ $\bullet$]  for the fermion
\begin{equation}
  \label{e:fermiontri12}
\hskip-2.2cm t_{12}^\chi(t,s)=  \frac{t}{8}\, \left(s^3 +3M^4 \left(s-u\right)-2s^2 u +2s u^2-u^3-2M^2\left(s^2-u^2\right)\right)\,.
\end{equation}
\end{itemize}

\noindent  {\it The massive triangle coefficients are given by }
\begin{itemize}
\item[ $\bullet$] for the scalar
\begin{multline}
  \label{e:scalartri34}
  t_{34}^\varphi(t,s)=\frac{1 }{4\left(t-4
      M^2\right)^2}\Big(-12 M^{10}+6 M^8 t-12 M^6 \left(s^2-3 s u+u^2\right)\cr +M^4 t \left(23 s^2-44 s u+23 u^2\right)-3
   M^2 t^2 \left(3 s^2-4 s u+3 u^2\right)+t^3 \left(s^2-s u+u^2\right) \Big)\,,
\end{multline}

\item[ $\bullet$] for the photon
\begin{multline}
  \label{e:photontri34}
  t_{34}^\gamma(t,s)=\frac{1 }{8 \left(t-4 M^2\right)^2}\,\Big(-4 M^{12} (s+u)+4 M^{10} (s-u)^2-16 M^8 (s-u)^2 (s+u)\cr -12 M^6 s u (s-u)^2+6 M^4 (s+u)
   \left(s^4+s^3 u-2 s^2 u^2+s u^3+u^4\right)\cr-M^2 (s-u)^2 (s+u)^2 \left(s^2-4 s u+u^2\right)-(s+u)^3 \left(s^2+u^2\right)
   \left(s^2-s u+u^2\right)\Big)\,,
\end{multline}
\item[ $\bullet$] for the fermion
\begin{multline}
  \label{e:fermiontri34}
  t_{34}^\chi(t,s)=\frac{2M^2-t-2s }{8 \left(t-4 M^2\right)^2}\,\Big(60M^{10} -2M^8(61t+60s)-t^3(t^2+3t s+3s^2)\cr+6 M^2 t^2(2t^2+6t s+5 s^2)\Big)\,.
\end{multline}
\end{itemize}

\break\noindent  {\it The massless bubble coefficients are}
\begin{itemize}
\item[$\bullet$] for the scalar
\begin{multline}
  \label{e:scalarbubble}
  bu^\varphi(t)=\frac{1  }{120(t-4
      M^2)^2}\,
\Big(-56 M^8+72 M^6 t+M^4 \left(23 s^2+10 s u+23
   u^2\right)\cr-M^2 t \left(13 s^2+218 s u+13 u^2\right)+t^2 \left(s^2+41
   s u+u^2\right)\Big)\,,
\end{multline}
\item[$\bullet$] for the photon
\begin{multline}
  \label{e:photonbubble}
  bu^\gamma(t)={su-M^4\over240
   \left(t-4 M^2\right)^2}\,\Big(1288 M^8-8 M^6 (373 t+322
   s)+M^4 \left(2083 t^2+3416 t s+1288 s^2\right)\cr
-2 M^2 t   \left(300 t^2+695 t s+532 s^2\right)+t^2 \left(60 t^2+163 t s+163 s^2\right) \Big)\,,
\end{multline}
\item[$\bullet$] for the fermion
\begin{multline}
  \label{e:fermionbubble}
  bu^\chi(t)={124 M^4 -272 M^2 t+49 t^2\over240
   \left(t-4 M^2\right)^2}\,\Big(2M^6 +2M^2 s(2t+3s)-M^4(t+6s)\cr-s(t^2+3t s +2s^2) \Big)\,.
\end{multline}
\end{itemize}

\section{The low energy limit of the cut constructible one-loop amplitude}
In the previous section we have provided the full non-analytic contributions to the
one-loop amplitudes of gravitational scattering of scalars, photons and fermions from
a large scalar mass. However, for many applications and specifically
for the calculation done in this paper,
only the leading low energy limit is
needed, and we present it here. In the low energy limit, the energy $E=\hbar \omega$ of
the massless particle is much smaller than the mass of the massive scalar
$\omega\ll M$ and the momentum transfer $t\sim -{\boldsymbol q}^2\ll M^2$ is tiny as well.

In this limit the one-graviton-exchange amplitude of a massless particle $X$
from the massive scalar $\Phi$ is given by  ({\it cf.}
appendix~\ref{sec:treegrav})
\begin{equation}
  \cM^{(1)}_X\simeq {\cN^X\over \hbar}\, \kappa^2 {(2M\omega)^2\over 4t}\,,
\end{equation}
where $\cN^{\varphi}=1$ for the massless scalar, while for
the photon $\cN^{\gamma}=(2M\omega)^{2} / (2\langle
p_1|p_3|p_2]^2)$ for the $(+-)$ photon helicity contribution and its complex
conjugate for the $(-+)$ photon helicity contribution, and
$\cN^{\chi}=M \omega/\langle p_1|p_3|p_2]$  for the $(+-)$ fermion helicity contribution and its
complex conjugate for the $(-+)$ configuration.  That the photon amplitude vanishes for the polarization configurations $(++)$ and $(--)$ is a direct consequence of the properties of the tree-amplitudes in Eq.~(\ref{e:ggGGhel}). Note that $|\cN^{\gamma}|^2\to1$ and $|\cN^{\chi}|^2\to1$ in the low-energy limit and therefore this pre-factor does not affect the cross-section.

The corresponding low energy one-loop amplitudes have the form
 \begin{multline}\label{e:OneL}
 \cM_{X}^{(2)}
 \simeq-{\cN^{X}\over\hbar}\, \bigg[
\hbar {\kappa^4\over 4} \, \Big(
4(M\omega)^4 (I_4(t,u)+I_4(t,s))+ 3(M\omega)^2 t I_3(t)
\cr-15 (M^2\omega)^2
I_3(t,M)+ bu^{X} (M\omega)^2 I_2(t)
 \Big)\bigg]\,,
 \end{multline}
where the coefficients of the bubble contributions are
\begin{equation}
bu^{\varphi}= {3\over40}\,, \qquad bu^{\gamma}=-{161\over 120}\,,\qquad
bu^{\chi}=-{31\over30}\,.
\end{equation}
Since in this limit $t\sim -{\boldsymbol q}^2$, $s\simeq (M+E)^2$, and $u\simeq
(M-E)^2+{\boldsymbol q}^2$, with $E=\hbar \omega$, the finite component of the integral functions $I_n$ are found to be ({\it cf.} appendix~\ref{sec:integral} for
more details)
\begin{eqnarray}
\label{e:boxeslog} I_4(t,s)+I_4(t,u)&\simeq& {1\over 2tM E}\, (4\pi)\,
  \log\left(-t\over M^2\right)  \\
\label{e:trimassless} I_3(t)&\simeq& -{1\over2t}\, \log^2\left(-t\over \mu^2\right)\\
I_3(t,M^2)&\simeq & -{1\over32\pi^2M^2}\, \left(\log\left(-t\over M^2\right)+
  {\pi^2M\over\sqrt{-t}}\right)\\
I_2(t)&\simeq&-{1\over 16\pi^2}\, \left(2-\log\left(-t\over\mu^2\right)\right)\,.
\end{eqnarray}
The total gravitational scattering amplitude
\begin{equation}
  \cM_X = {1\over \hbar} \cM_X^{(1)}+  \cM_X^{(2)}\,,
\end{equation}
then has the low-energy expansion 
\begin{eqnarray} \label{e:result1}
&&   
 \cM_X^{(2)}
 \simeq{\cN^{X}\over \hbar}\,  (M\omega)^2\cr
&\times&
 \Big[-{\kappa^2\over \boldsymbol q^2}-\kappa^4 {15\over
  512}{ M\over |\boldsymbol q|}
-\hbar \kappa^4
 {15\over
  512\pi^2}\,\log\left(\boldsymbol q^2\over M^2\right)
+\hbar\kappa^4\,{ bu^{X} \over(8\pi)^2} \, \log\left(\boldsymbol q^2\over \mu^2\right)
\\
\nonumber &-& \hbar\kappa^4 {3\over128\pi^2}\, \log^2\left(\boldsymbol q^2\over \mu^2 \right)
+\kappa^4 \,  {M\omega\over 8\pi} {i\over \boldsymbol q^2}\log\left(\boldsymbol q^2\over
  M^2\right)
\Big]\,,
\end{eqnarray}
where $\mu^2$ is a mass scale parameter used in dimensional
regularization. This result confirms ref.~\cite{Bjerrum-Bohr:2014zsa} and extends it to the case of
fermonic scattering. Note that Eq.~(\ref{e:result1}) contains both
classical (independent of $\hbar$) and quantum mechanical ($\propto
\hbar$) loop contributions.  It was shown in~\cite{Holstein:2004dn}
why classical post-Newtonian corrections appear in one-loop
gravitational amplitudes.  While most field theories identify
classical and quantum effects by separating tree from loop topologies,
this is not a fundamental distinction and has in fact more to do with
linearity vs. non-linearity of the field equations; it is then natural
that a quantum field theory constructed from the non-linear
Einstein-Hilbert action receives classical contributions from loop
topologies.  At one-loop order, in the above computation, this
contribution is of the type $\sim\kappa^4/ |\boldsymbol q|$ and it is
a very pleasing check of our computation that we observe universality in both particle type and statistics for this contribution, as expected from general relativity.

The one-loop amplitude has infrared divergences
arising from the propagation of the graviton between the
massless external legs $p_1^2=p_2^2=0$,
%\begin{figure}[h]
 % \centering
\begin{fmffile}{IRdiv}
\begin{equation}   \label{fig:IR}
\begin{gathered}\begin{fmfgraph*}(110,80)
      \fmfleftn{i}{2}
      \fmfrightn{o}{2}
	\fmfrpolyn{smooth,filled=30}{G}{4}
\fmf{dashes_arrow,label=$p_1$}{i1,v1}
\fmf{dashes_arrow,label=$p_2$}{i2,v2}
\fmf{dashes,tension=.8,label=$\ell$,label.side=left}{v1,v2}
\fmf{plain,tension=2.5}{G2,o1}
\fmf{plain,tension=2.5}{G1,o2}
\fmf{dbl_curly,tension=.6}{G3,v1}
\fmf{dbl_curly,tension=.6}{G4,v2}
\end{fmfgraph*}
\end{gathered}
\propto \cM^{(1)}_X\times\int_0 {d^{4-2\epsilon}\ell\,\mu^{2\epsilon}\over \ell^2\, 2\ell\cdot p_1\, 2\ell\cdot
  p_2}\sim {(t\,\mu^{-2})^{-\epsilon}\over t\,\epsilon^2}\,\cM^{(1)}_X\,.
\end{equation}
\end{fmffile}
%  \caption{The one-loop IR divergence arise }
%\end{figure}
%
The resulting infrared  divergence is
contained in the scalar boxes and the massless triangle in the
decomposition~\eqref{e:OneL}.  The infrared divergences of
gravitational theories have been studied in~\cite{Weinberg:1965nx,Naculich:2011ry,White:2011yy,Akhoury:2011kq,Donoghue:1999qh}. At
one-loop  the amplitude is the sum,
$\cM_X^{(2)}=  \mathcal S\, \cM^{(1)}+  \mathcal H$, consisting
of an infrared divergent part from the soft region in~\eqref{fig:IR},
$\mathcal S=(t \mu^{-2})^{-\epsilon}/(t \epsilon^2)$ times $\cM^{(1)}$ (the
tree-level one-graviton amplitude is evaluated
in appendix~\ref{sec:treegrav})
and a finite hard part $\mathcal H$.
When forming cross-sections, we know how to resolve them. Soft graviton bremsstrahlung radiation
also contributes to the measured cross-section if the radiated graviton is below the
resolution of the detector. Including the cross-section for bremsstrahlung with a finite detector
resolution $\Delta E$ has the effect of converting the scale $\mu$ into the detector resolution,
potentially along with some finite constants depending on the specifics of the detector and
the cross-section definition. This has been
checked explicitly for massless gravitons in the
process of graviton-graviton scattering~\cite{Donoghue:1999qh}.

In our case, we are about to use this amplitude in
the process of light bending. Again, very soft gravitational bremsstrahlung cannot be
differentiated from the non-radiative light bending, and should be included in the measurement.
As with the cross-section, this should eliminate the IR divergences and replace the scale $\mu$
the logarithm by a factor depending on the resolution of the measurement. We have not done an explicit
calculation of this process. If
the quantum correction were close to being observable and if detectors capable of resolving
graviton bremsstrahlung existed, one would be motivated to perform a careful analysis. However, a couple of
options present themselves. If the light was a monochromatic beam, the detector resolution could be a
resolution in energy of the photon. This could be either an energy independent resolution factor, or one
which is proportional to some fraction of the original energy. These two cases would then have different infrared
factors in the bending angle, indicating that there is not a unique
detector-independent factor to include in a bending angle formula. Alternatively, the angular resolution of the detector could be used to define the acceptance
factor for graviton bremsstrahlung. In the absence of a full calculation, we simply replace the scale $\mu$ in the logarithm by
and infrared scale which we will call $1/b_0$ below.

%%%%%%%%%%%%%%%%%%%%%%%%%%%%%%%%%%%%%%%%%%%%%%%%%%%%%%%%%%%%%%%%%
\section{Bending of light}
\subsection{Bending formula from general relativity}

Perhaps the most famous verification of Einstein's general theory of relativity is its prediction for the bending angle of light passing the rim of the Sun, since its 1919 measurement during a total solar eclipse led to worldwide publicity and acceptance of Einstein's theory.  The standard derivation of this result in general relativity follows from considering a spherically symmetric metric parameterized as
\begin{equation}\label{e:QSch}
    ds^2= A(r) dt^2 - B(r)^2 dr^2 - r^2 d\Omega^2\,.
\end{equation}
In the case of the Schwarzschild metric we have
\begin{equation}
A(r)={1\over B(r)}=1-{2G_NM\over r}\,,
\end{equation}
so that geometrically the deflection angle is given the standard formula
\begin{equation}
\theta=2\int_0^1{du\over \sqrt{1-u^2-{2G_NM\over R}(1-u^3)}}-\pi\,,\label{eq:bv}
\end{equation}
where we have defined $u={R\over r}$.  Here $R$ is the distance of closest approach in Scwarzschild coordinates.  The integration in Eq. (\ref{eq:bv}) can be performed exactly in terms of elliptic functions, but since near the solar rim $2G_NM/R\simeq 10^{-3}\ll1$, we can instead use a perturbative solution
\begin{eqnarray}
\theta&=&2\int_0^1du\left[{1\over \sqrt{(1-u)(1+u)}}+{G_NM\over R}{1+u+u^2\over \sqrt{(1-u)(1+u)^3}}\right.\nonumber\\
&+&\left.{3\over 2}{G_N^2M^2\over R^2}{(1+u+u^2)^2\over \sqrt{(1-u)(1+u)^5}}+\ldots\right]-\pi\nonumber\\
&=&{4G_NM\over R}+{4G_N^2M^2\over R^2}\left({15\pi\over 16}-1\right)+\ldots
\end{eqnarray}
However, instead of using the coordinate-dependent quantity $R$, the bending angle should be written in terms of the impact parameter $b$, defined as
\begin{equation}
b=\sqrt{B(R)}R={R\over \sqrt{1-{2G_NM\over R}}}=R+G_NM+\ldots
\end{equation}
It is important to note that $b$ is a coordinate-{\it  independent} quantity whereas $R$
depends on the coordinate system (see~\cite{Willangle} for a nice
discussion of the coordinate dependence on the expression of the
deflection angle).  We arrive then at the universal (matter-independent) formula for the bending angle
\begin{equation}
  \label{e:rho2GR}
{b\over 2G_NM}= {2\over\theta}+ {15\pi\over32}+ \mathcal O(\theta)\,,
\end{equation}
or
\begin{equation}
\theta={4G_NM\over b}+{15\pi G_N^2M^2\over 4b^2}+{\cal O}\left({1\over b^3}\right)\,.
\end{equation}
This is the standard derivation and arises from considering light as particles (photons) traversing a classical trajectory.  Below we show how we can reproduce this expansion from the (low energy) limit of the one-loop scattering amplitude, which can be thought of as a quantum mechanical (wavelike) derivation.

\subsection{Leading Newtonian correction}
We will first reproduce the leading term by evaluating the (classical) elastic differential
cross-section using only the first Newtonian (tree-level) contribution. Writing the cross-section out as
a perturbative expansion we have
\begin{equation}
\cM=\cM^{(1)}+\cM^{(2)}+\cdots
\end{equation} and  thus
\begin{equation}
|\cM|^2= |\cM^{(1)}|^2+ 2 \Ree( \cM^{(1)} (\cM^{(2)})^*)+\cdots
\end{equation}
Since we are interested in the low energy limit $E\ll m$, we can make the
approximations $t=(p_1+p_2)^2=(p_3+p_4)^2\simeq -{\boldsymbol q}^{2}=-4
E^2 \,(\sin \theta/2)^2$, with $q=p_1+p_2$,
$s=(p_1+p_4)^2\simeq (M+E)^2\simeq M^2+2ME$, and $u=(p_1+p_3)^2\simeq
(M-E)^2+{\boldsymbol q}^2\simeq M^2-2ME+{\boldsymbol q}^2$. We also employ a
small angle scattering approximation so that $|{\boldsymbol q}^2|\ll E^2$.
Recalling that $\kappa^2=32\pi G_N/c^4$, the tree-level contribution to the cross-section is given by
\begin{equation}
  \left. d\sigma\over dt  \right|^{\rm tree}_{\varphi^2}=\frac{\kappa^4 (u-M^2)^2}{16^3 \pi t^2} \simeq \pi\,\left( 4 G_N M    E\over t\right)^2\,.
\end{equation}
Making the (classical) assumption that we can determine the impact parameter $\rho(\theta)$ from the cross-section $d\sigma=\pi d\rho^2$, we have
\begin{equation}
  \rho^2 =\int_{4E^2(\sin\theta/2)^2}^{4E^2} {d\sigma\over dt} \,
  {dt\over \pi}\,,
\end{equation}
and in the small angle approximation $\theta\ll1$
\begin{eqnarray}
  \rho^2&\simeq&
  (4 G_N ME)^2 \int_{4E^2(\theta/2)^2}^{4E^2} \, {dt\over t^2}\simeq
\left(4G_N M\over \theta\right)^2\,,
\end{eqnarray}
which gives the relation between the bending angle and the impact parameter
\begin{equation}
  \theta\simeq {4G_NM\over \rho}\,.
\end{equation}
It is clear, given Eq.~(\ref{e:result1}), that this result for the leading contribution to the bending angle is universal for both particle type and statistics and agrees with general relativity.
Of course, the validity of this semi-classical derivation of the leading contribution (Coulomb-type potential scattering) to the bending angle is {\it not} guaranteed when considered within a full quantum mechanical framework. It is, however, still true, up to a phase, due to the fact that even quantum mechanically angular momentum remains conserved for a Coulombic potential, so that the classical cross-section formula at leading order is valid even in the full quantum regime.

\subsection{Bending via the Eikonal Approximation}

An appropriate quantum mechanical treatment of the light bending
problem makes use of the eikonal formulation, which describes the
scattering in terms of an impact parameter representation. In impact
parameter space, the scattering amplitude exponentiates into an
eikonal phase and evaluation using the stationary phase method yields
the classical result for the bending angle, together with quantum  effects.

There are two important aspects to the eikonal approximation. One is kinematic. When the impact parameter is large, the bending angle is small. The small-angle approximation means that the momentum transfer is small---$t\sim -{\boldsymbol q}^2\ll s$. This condition is easy to implement in our amplitude.  The second approximation is diagrammatic. The leading eikonal approximation involves the iteration of one graviton exchange in all permutations. The first correction to this approximation involves more complicated diagrams, such as loop processes, in addition to the permutations
of graviton exchange. The leading eikonal phase is of order $G_N$, and
the first correction to the phase will be of order $G_N^2$. We will
impose eikonal kinematics and proceed to the first diagrammatic
correction.

Our guide in this approach is the discussion of the next-to-leading
eikonal amplitude by Akhoury et al. in~\cite{Akhoury:2013yua}\footnote{ We note
that, at the time our manuscript is being written, the preprint
version of their work contains a clear error in the summary of their
amplitude. We have confirmed this with the authors and a
corrected version of this work will appear soon. Once this mistake is rectified, their result reproduces the correct next-to-leading classical bending angle.}. This method can be readily generalized to include the quantum terms within the same diagrams as well as the purely quantum diagrams (the bubble diagrams) at the same order. We determine the eikonal phase by matching the amplitudes at one-loop order.
In high energy small-angle scattering the dominant four-momentum
transfer is in the transverse spatial direction. For photons traveling
in the $z$ direction we have $p_3=p_1 +q$ so that, squaring this
equation, we obtain $0=2E(q_0 -q_z) +\boldsymbol q^2$. A similar
calculation for the heavy scalar yields $0=-2Mq_0 +\boldsymbol q^2$,
which tells us that both $q_\pm = q_0 \pm q_z$ are suppressed compared
to the transverse components $\boldsymbol q^2\sim -\boldsymbol q^2_\perp$ by at least a factor of $2E$. This condition on the overall momentum transfer gets reflected in the same condition on the exchanged gravitons, so that the dominant momentum transfer inside loops is also transverse. In the effective theory of high energy scattering, the Soft Collinear Effective Theory (SCET), these are called Glauber modes~\cite{Rothstein:2016bsq} and carry momentum scaling $(k_+,k_-,k_\bot)\sim \sqrt{s}(\lambda^2,\lambda^2,\lambda)$ where $\lambda \sim \sqrt{-t/s}$.

The one-graviton amplitude amplitude in this limit is 
\begin{equation}
{\cal M}^{(1)}_1(\boldsymbol q) = \kappa^2M^2E^2
\frac{1}{{\boldsymbol q}^2}\,,
\end{equation}
and, after some manipulations, the multiple exchanges of this amplitude can be arranged into a form which exponentiates
\begin{equation}
{\cal M}^{(1)}_{sum}(\boldsymbol q) =32\pi ME   \sum_n \frac{1}{n!} \left(2G_NME\right)^n\prod_{i=1}^n \int \frac{d^2k_i}{(2\pi)^2}\frac{1}{\mathbf{k}_i^2} \delta^2(\sum k_i-q)\,.
\end{equation}
In order to bring this amplitude into impact parameter space, one defines the Fourier transform, with
impact parameter ${\boldsymbol b}$ being transverse to the initial motion.
\begin{equation}
{\cal M}({\boldsymbol b}) =  \int \frac{d^2\boldsymbol q}{(2\pi)^2}
e^{-i{\boldsymbol q}\cdot {\boldsymbol b}}~ {\cal M}^{(1)}_{sum}(\boldsymbol q)\,.
\end{equation}
Writing the sum as an exponential, the result (with the prefactor relevant for gravity) for the scattering of a massless particle from a massive one is given by
\begin{align}
{\cal M}({\boldsymbol b}) = 2(s-M^2) \left( e^{i\chi_1({\boldsymbol b})} -1\right)\,.
\end{align}
Here $\chi_1({\boldsymbol b})$ is the Fourier transform of the one graviton exchange, with some kinematics factored out
\begin{equation}
\chi_1({\boldsymbol b}) = \frac{1}{2M2E}\int
  \frac{d^2\boldsymbol q} {(2\pi)^2}~e^{-i{\boldsymbol q}\cdot {\boldsymbol b}}~ {\cal
    M}_1(\boldsymbol q) \,,
\end{equation}
and can be evaluated using dimensional regularization,
\begin{equation}
  \int {d^d \boldsymbol q\over(2\pi)^2}  \, e^{-i\boldsymbol q\cdot \boldsymbol
    b} \, |\boldsymbol q|^\alpha =\int_0^\infty {dq\over 2\pi}
  \, q^{\alpha+d-1}
  J_0(q b) = {1\over4\pi (2b)^{\alpha+d}}\,
  {\Gamma\left(\alpha+d\over2\right)\over\Gamma\left(2-\alpha-d\over2\right)}\,,
\end{equation}
and taking the limit $d\to2$ in the final expression.
Therefore (using $\kappa^2=32\pi G_N$)
\begin{align}
\chi_1(\boldsymbol b)
&= {\kappa^2ME\over4}\int \frac{d^2 \boldsymbol q}{(2\pi)^2}~e^{-i{\boldsymbol q}\cdot
  {\boldsymbol b}} \frac{1}{{\boldsymbol q}^2}\ \ \nonumber \\
  &\simeq 4G_NME\left[\frac{1}{d-2} -\log (b/2)-\gamma_E\right] \,,
\end{align}
with $E$ being the energy of the massless particle. Only the $\log b$ term will be important in the following treatment.

At order $G_N^2$, the matrix element picks up corrections which we can describe by
\begin{equation}\label{e:Mtot}
{\cal M}(\boldsymbol q) = {\cal M}^{(1)}_1(\boldsymbol q)+ {\cal M}^{(2)}(\boldsymbol q)\,,
\end{equation}
where ${\cal M}^{(2)}(\boldsymbol q)$ is our calculated amplitude evaluated in this kinematic limit.
Including the dressing of ${\cal M}^{(2)}(\boldsymbol q)$ by permutations of one
graviton exchange, it was shown in~\cite{Akhoury:2013yua} that there is again an exponentiation of the simple exchange
\begin{equation}
{\cal M}^{(2)}_{sum}(\boldsymbol q) =(32\pi)^2 ME   \sum_n \frac{\left(2G_NME\right)^{n-2}}{(n-2)!} \int d^2k_j \cM^{(2)}(k_j)\prod_{i\ne j}^n \int \frac{d^2k_i}{(2\pi)^2}\frac{1}{\mathbf{k}_i^2}
\delta^2(\sum k_i-q)\,.
\end{equation}
In impact parameter space then we can write~\eqref{e:Mtot}
\begin{equation}
{\cal M}(\boldsymbol b) = 2 (s-M^2)  \left[(1+i\chi_2) e^{i\chi_1} -1 \right]\simeq 2 (s-M^2)  \left[e^{i(\chi_1+\chi_2)} -1 \right]\,,
\end{equation}
with the second expression being valid to this order in $G_N$.
The second order phase $\chi_2$ is given by
\begin{equation}
\chi_2({\boldsymbol b}) = \frac{1}{2M2E}\int
\frac{d^2\boldsymbol q}{(2\pi)^2}~e^{-i \boldsymbol q\cdot {\boldsymbol b}}~ {\cal M}^{(2)}(\boldsymbol q) \,.
\end{equation}
For the classical correction we need the integral
\begin{equation}\label{classicalint}
  \int{d^2\boldsymbol q\over (2\pi)^2}e^{-i\boldsymbol{q}\cdot\boldsymbol{b}}{1\over |\boldsymbol{q}|}=\frac{1}{2\pi b}\,,
\end{equation}
while for the quantum terms we require
\begin{eqnarray}\label{quantint1}
  \int{d^2q\over
  (2\pi)^2}e^{-i\boldsymbol{q}\cdot\boldsymbol{b}}\log\boldsymbol{q}^2&=&-{1\over
  \pi b^2}\,,\\
\label{quantint2}
  \int{d^2q\over (2\pi)^2}e^{-i\boldsymbol{q}\cdot\boldsymbol{b}}\log^2\boldsymbol{q}^2&=&{4\over \pi b^2}\log{2\over b}\,.
\end{eqnarray}
We find then
\begin{equation}\label{chi2}
  \chi_2 (\boldsymbol{b}) = G_N^2M^2E\frac{15\pi}{4b} + \frac{G_N^2M^2E }{2\pi b^2}\left(8 bu^\eta -15 + 48 \log \frac{2b_0}{b} \right)\,.
\end{equation}

The light bending analysis is now straightforward and involves determining the stationary phase of the exponent, which can be argued to dominate the momentum space integration, via
\begin{equation}
\frac{\partial}{\partial b} \left(-\boldsymbol q \cdot \boldsymbol b +
  \chi_1(b)+\chi_2(b)+\cdots\right)=\frac{\partial}{\partial b}
\left(q\, b + \chi_1(b)+\chi_2(b)+\cdots\right) =0\,.
\end{equation}
Using $q =2E \sin (\theta/2)$ this condition reads
\begin{equation}
2 \sin  \frac{\theta}{2} \simeq \theta =-\frac{1}{E}\frac{\partial}{\partial b} \left( \chi_1(b)+\chi_2(b)\right) \,,
\end{equation}
and yields
\begin{equation}\label{e:theta}
\theta  \simeq {4G_N M\over b}+{15\over 4} {G_N^2 M^2 \pi \over
  b^2}
+\left(8 bu^S+9-48 \log {b\over 2b_0}\right)\,{\hbar G_N^2 M\over  \pi b^3}+\ldots\,.
\end{equation}
 Here $1/b_0$ in the logarithm is the infrared cutoff
which removes the IR singularities of the amplitude.
We see that the eikonal  approximation leads to the expected
classical  general relativity contributions, in agreement  with
 the next-to-leading
correction of~\cite{Akhoury:2013yua} and \cite{D'Appollonio:2010ae}, as well as producing the leading quantum correction.
Treating the quantum
effect using the eikonal procedure,  we recover the results of~\cite{Bjerrum-Bohr:2014zsa} derived
with a semiclassical potential method.

The quantum effect has the power-law  dependence in impact
parameter as  the  classical
  post-post-Newtonian contribution.  This second post-Newtonian
  contribution of order
  $G_N^3M^3/b^3$  arises as a classical piece from two-loop
amplitudes with  momentum dependence $\boldsymbol
q^0/\hbar$. These two contributions lead to  very distinct
analytic structure to the $S$-matrix   and are easily separated.  The
classical corrections is much larger than the quantum effect by the
ratio the square of the Schwarzschild radius to the Planck length
$G_N^2M^2\gg \hbar G_N$.

\subsection{Bending via Geometrical Optics}

There is an equivalence between the eikonal method described above and the semiclassical potential method which we used in \cite{Bjerrum-Bohr:2014zsa}. In order to elucidate this, it is useful to consider the bending in terms of a wave picture of light propagation.  Since the wavelength of the light is much smaller than size of the massive scalar object (Sun or black hole) around which the bending occurs, the analysis can be done using the methods of geometrical (ray) optics.  This formalism is developed in many places, {\it e.g.}~\cite{brau}, and leads to the equation
\begin{equation}
{d\over ds}n{d\boldsymbol{r}\over ds}=\boldsymbol{\nabla}n\,,\label{eq:jm}
\end{equation}
where $n$ is the index of refraction and $\boldsymbol{r}(s)$ is the trajectory as a function of the path length $s$.  For light we can write $ds\simeq cdt$ so that Eq.~(\ref{eq:jm}) becomes
\begin{equation}
{1\over c^2}{d^2\boldsymbol{r}\over dt^2}={1\over n}\boldsymbol{\nabla}n\,.
\end{equation}
In our case, at leading order, the index of refraction is determined from the general
relativity/optical-mechanical analogy~\cite{Alsing} which, for a line element
\begin{equation}
ds^2=A(r)dt^2-B(r)\,dr^2+r^2(d\theta^2+\sin^2\theta d\phi^2)\,,
\end{equation}
yields
\begin{equation}
n(r)=\sqrt{B(r)\over A(r)}\,.
\end{equation}
For the Schwarzschild metric we have then
\begin{equation}
n(r)={1\over 1-{2G_NM\over r}}\simeq 1+{2G_NM\over r}=1-{1\over E_m}V_0(r)\,,
\end{equation}
where
\begin{equation}
V_0(r)=-{2G_NME_m\over r}\,,
\end{equation}
is the leading order potential energy for a photon of energy $E_m$ interacting with a massive scalar of mass $M$. Following~\cite{Bjerrum-Bohr:2014zsa}, the generalization to the full interaction is
then achieved by replacing the lowest order potential $V_0(r)$ by the full interaction potential $V_{\rm int}(r)$ generated from the low energy approximation with $t\simeq -\boldsymbol q^2$ of the
total gravitational scattering amplitude\footnote{We made use of the following Fourier
  transformations
  \begin{equation}
    \int {d^3{\boldsymbol q}\over(2\pi)^3} \, e^{i\boldsymbol q\cdot \boldsymbol r}
    {1\over \boldsymbol q^2}= \frac{1}{4\pi r}\,, \qquad
 \int {d^3{\boldsymbol q}\over(2\pi)^3} \, e^{i\boldsymbol q\cdot \boldsymbol r}
    {1\over |\boldsymbol q|}= {1\over 2\pi^2 r^2}\,,\qquad
 \int {d^3{\boldsymbol q}\over(2\pi)^3} \, e^{i\boldsymbol q\cdot \boldsymbol r}
    \log (\boldsymbol q^2)= -{1\over 2\pi r^3}\,.
\end{equation}
} %
in Eq.~(\ref{e:result1}) with the second Born (rescattering) term excised
\begin{eqnarray}\label{e:V}
  V_{\rm int}( r)&=& {\hbar\over 4M\omega} \int \left(\cM_X(\boldsymbol q)+\kappa^4{M\omega\over 8\pi\boldsymbol{q}^2}\log{\boldsymbol{q}^2\over M^2} \right)\, e^{i \boldsymbol
    q\cdot \boldsymbol r} {d^3\boldsymbol q\over (2\pi)^3}\\
\nonumber&=&-{2G_N M E\over r}-{15\over4} {(G_NM)^2E\over r^2}-
            {8bu^S-15+48\log(r/r_0)\over 4\pi}\, {\hbar G_N^2ME\over r^3}\,.\label{eq:gj}
\end{eqnarray}
(Note that this interaction potential is {\it not} derivable from an quantum corrected effective  background metric. We will comment more about this fact in the section~\ref{sec:conclusion}.)
In the absence of a potential, we can imagine a photon incident in the $\hat{e}_y$ direction with impact parameter $b$ on a massive scalar target located at the origin.  The trajectory is then characterized by
\begin{equation}
\boldsymbol{r}(t)=b\hat{e}_x+ct\hat{e}_y,\qquad -\infty<t<\infty \ .
\end{equation}
If we now impose a potential, there will exist a small deviation from this straight line trajectory with
\begin{equation}
\Delta {1\over c^2}{d\boldsymbol{r}\over dt}=-\int_{-\infty}^{+\infty}
dtV_{\rm int}'(r)\hat{\boldsymbol{r}}\,.
\end{equation}
We have then
\begin{equation}
{2\over c}\sin{1\over 2}\theta\simeq {1\over
  E_m}\int_{-\infty}^{+\infty} dtV_{\rm int}'(\sqrt{b^2+c^2t^2}){b\over
  \sqrt{b^2+c^2t^2}}\,,
\end{equation}
where $\theta$ is the scattering angle.  Changing variables to $t=bu/c$, we find
\begin{equation}
{2\over c}\sin{1\over 2}\theta\simeq {1\over c}\theta=-{b\over
  E_m}\int_{-\infty}^{+\infty} V_{\rm int}'(b\sqrt{1+u^2}){dux\over
  \sqrt{1+u^2}}\,,
\end{equation}
yielding
\begin{equation}
\theta={b\over E_m}\int_{-\infty}^{+\infty} V_{\rm int}'(b\sqrt{1+u^2}) {du\over
  \sqrt{1+u^2}}\,.
\end{equation}
Substituting the interaction potential Eq.~(\ref{eq:gj}) and performing the requisite integration, we arrive at
\begin{equation}
\theta = {4G_N M\over b}+{15\pi G_N^2M^2\over 4b^2}+
\left(8 bu^S+9-48 \log {b\over 2b_0}\right)\,{\hbar G_N^2M\over \pi b^3}+\ldots\,.\label{eq:pk}
\end{equation}
The first two (classical) pieces of Eq.~(\ref{eq:pk}) agree with the standard post-Newtonian analysis given above, but they are accompanied by a small quantum mechanical correction term found in our eikonal analysis. It may seem surprising that small angle scattering theory, involving the two-dimensional eikonal Fourier transform yields a result identical to that found from the semiclassical potential result, which is given in terms of a three-dimensional Fourier transform of the transition amplitude.
However, this equality is made clear from the mathematical identity
\begin{equation}
\int_0^\infty dqq^2J_1(qb)F(\boldsymbol q^2)={b\over \pi}\int_{-\infty}^\infty {du\over \sqrt{1+u^2}}\int_0^\infty dq q^3j_1(qb\sqrt{1+u^2})F(\boldsymbol q^2)
\end{equation}
which is valid for any sufficiently smooth function $F(\boldsymbol q^2)$~\cite{Gra65}.

%%%%%%%%%%%%%%%%%%%%%%%%%%%%%%%%%%%%%%%%%%%%%%%%%%%%%%%%%%%%%%%%%%%
\section{Discussion}\label{sec:conclusion}
%%%%%%%%%%%%%%start over%%%%%%%%%%%%%%%%%%%%%

We have in this presentation derived one-loop scattering results for all types of massless matter interacting gravitationally with a massive scalar source. While we have found universality and agreement with general relativity for the classical physics component of the result, {\it i.e.}, the so called post-Newtonian corrections, field theory has also produced a new type of contribution
of quantum origin, which has no precedence in classical general
relativity. This pattern of new contributions will persist to all loop orders, and thus produce terms having varying powers of $\hbar$ that all are unique signatures of quantum effects in the theory of gravity. We will here comment on
on the role of such terms.

While many field theories have limits wherein quantum effects can be dealt with and motivated in a semi-classical/semi-quantum context (even in QCD!) it is particularly hard finding such limits in
general relativity, given its geometric nature and local description. Concepts like a free falling elevator and motion along geodesics have no known simple quantum mechanical equivalent.
An interesting observation, however, is that such quantum terms, except for the bubble coefficients, are universal. The bubble non-universality could be interpreted as a violation of some classical descriptions of the
equivalence principle, in that massless particles do not follow null geodesics, and different types of massless particles follow different trajectories. However this in not a fundamental violation of the equivalence principle in the larger sense, as the action which defines the theory is compatible with the equivalence principle. However, in the scattering process tidal effects offer another possible interpretation of the result, since we have the quantum loops of massless particles involving long-range propagation in a non-homogenous gravitational field. Construction of a gedanken experiment featuring a homogeneous gravitational field could thus be an interesting exercise.

One might consider the possibility that the quantum correction to the gravitational interaction between two massive  particles could have a geometrical interpretation in terms of an effective particle evolving in a quantum-corrected metric.  However, this seems not to be feasible since the effective one-particle reducible potential that would result from propagation in a quantum corrected metric would be gauge-dependent.  This is already the case for the interaction potential between two massive particles. It appears that a fully quantum mechanical description, such as we have presented, is required.

We conclude that the best way to deal with this situation is to simply
compute a cross-section for scattering
and use this to compare observational data to theory. The effects are seen to be too small to
be observed experimentally, yet they can yield interesting theoretical
insights, such as the evidence that massless particles no longer
follow null geodesics, and that the cross-section is not universal as it
depends on the type of massless particle.

%--------------------------------------------------------------------------
\section*{Acknowledgement}

We thank Costas Bachas, Massimo Bianchi, Poul H. Damgaard, C\'edric
Deffayet, Gregory Korchemsky, Ugo Moschella for useful discussions and
George Sterman and Ratin Akhoury for useful communications. We thank
Thibault Damour for pointing out  incorrect signs and misplaced
factors of $i$ in a previous version
of this paper.
The research leading to these results has  in part been supported
by the Danish National Research Foundation (DNRF91) as well as the ANR
grant reference QST ANR 12 BS05 003 01, and the PICS 6430.
PV is partially supported by a fellowship funded by the French Government at Churchill College, Cambridge. The research of JFD and BRH is supported in part by the National Science Foundation under grants NSF PHY15-20292 and NSF PHY12-25915 respectively.
%--------------------------------------------------------------------------
\appendix

\section{Gravitational photon and scalar tree amplitudes}\label{sec:gravtree}

\subsection{Helicity formalism conventions}\label{sec:conventions}

This appendix contains a brief account of the  conventions and the
notation in the paper.
We  follows the notations and conventions of  ref.~\cite{Mangano:1990by}.

We employ the mostly minus metric signature
$\eta^{\mu\nu}=\textrm{diag}(+,-,-,-)$ and use Dirac matrices satisfying
$\{\gamma^\mu,\gamma^\nu\}=2\eta^{\mu\nu}$, {\it i.e.},
\begin{equation}
\gamma^\mu =
\begin{pmatrix}
  0 & \sigma^\mu \\
  \bar{\sigma}^\mu & 0
\end{pmatrix}; \qquad
\gamma_5=\begin{pmatrix}
1 & 0
\\ 0 & -1
\end{pmatrix}\,.
\end{equation}
We have  $(\sigma^\mu)=(1,\sigma^i)$ and
$(\bar\sigma^\mu)=(-1,\bar\sigma^i)$ where $\sigma^i$ are the standard
Pauli matrices. We use some places the notation $\gamma^\mu
p_\mu=\sla p$.
The Dirac matrices satisfy the Clifford algebra (we
refer to~\cite[App. A-2]{Itzykson:1980rh} for details)
$\{\gamma^\mu,\gamma^\nu\}=2\eta^{\mu\nu}$ and $\gamma_5$ is the
chirality operator, satisfying $
\tr(\gamma_5\gamma^{\mu_1}\gamma^{\mu_2}\gamma^{\mu_3}\gamma^{\mu_4})=4i\,\epsilon^{\mu\nu\rho\sigma}$.

We have the following conventions for traces. They are defined by $\tr_\pm(a_1 \cdots a_r ):=\tr\big(
{1\pm\gamma_5\over2} \gamma^{\mu_1}\cdots \gamma^{\mu_r}\big)
a_{1\,\mu_1}\cdots a_{r\,\mu_r}$. We note in
particular that $\tr_\pm(abcd)=2 (a\cdot b\, c\cdot d - a\cdot c \, b\cdot d+a\cdot d\,
  b\cdot c)\pm 2i \varepsilon^{\mu\nu\rho\sigma} a_\mu b_\nu
c_\rho d_\sigma$.

\medskip
 The Levi-Civita epsilon tensor 
$\varepsilon^{\mu\nu\rho\sigma}$ takes the value 1 if $\{\mu,\nu,\rho,\sigma\}$ is an
even permutation of $\{0,1,2,3\}$, $-1$  if $\{\mu,\nu,\rho,\sigma\}$ is an
odd permutation of $\{0,1,2,3\}$, and 0 otherwise.

\bigskip

For a light-like momentum $p$ the positive energy solution to the
Dirac equation is $\sla p  \, u_h(p)=0$ both for positive and
negative helicities, {\it i.e.}, $h=+1$ and $h=-1$. This solution
satisfy the usual chirality condition $(1\mp \gamma_5)/2\, u_\pm(p)=0$ and
$(1\pm\gamma_5)/2\,\bar u_\pm(p)=0$.

We will also make use of the following conventions
\begin{eqnarray}
|k\rangle{}&\equiv u_+(k);\qquad |k]{}&\equiv
u_-(k) \\
\langle k|{}&\equiv \bar u_-(k);\qquad {}[k|{}&\equiv \bar u_+(k)\,.
\end{eqnarray}
and spinor products will be defined according to
\begin{equation}
\an[p,q]{} \equiv\bar u_-(p) u_+(q);\qquad \sq[p,q]{}\equiv \bar
u_+(p) u_-(q)\,,
\end{equation}
where $(p+q)^2=2p\cdot q=\an[p,q]\sq[p,q]$.

This yields the following completeness relation
\begin{equation}
\sum_{h=\pm1} u_h(k)\bar u_h(k)=\sla k= |k \rangle[k|+ |k]\langle k|\,.
\end{equation}
and we arrive at 
\begin{equation}
\label{e:pol} 
\sla\epsilon^+_\mu(k,p_{\rm ref})
= { [ k | \gamma_\mu | p_{\rm ref} \ra
\over 
\sqrt2 \an[p_{\rm ref},k]};
\qquad 
\sla \epsilon^-_\mu(k,p_{\rm ref})
= -{ \langle k | \gamma_\mu | p_{\rm ref} ]
\over 
\sqrt2 \sq[p_{\rm ref},k]}
\,,
\end{equation}
for the polarisation tensor for the photon of light-like momentum $k$ where $p_{\rm ref}$ is an 
arbitrary light-like reference momentum.

For $p$ and $q$ light-like momenta and $k$ a four-momentum vector we have
\begin{equation}
    \langle p|k|q] = \langle p|\gamma^\mu|q] \, k_\mu; \qquad [p|kq\rangle=(  \langle p|k|q] )^*\,.
\end{equation}

\subsection{The gravitational Compton amplitudes}\label{sec:compton}

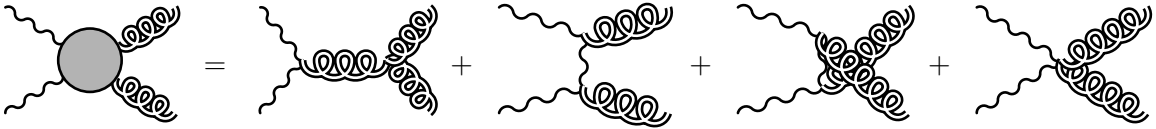
\begin{figure}[ht]
  \centering
\begin{fmffile}{figcompton}
\begin{equation*}
\begin{gathered}
    \begin{fmfgraph*}(80,40)
       \fmfleft{i1,i2}
        \fmfright{o1,o2}
\fmf{photon}{i1,v1}
\fmf{photon}{i2,v1}
\fmf{dbl_curly}{v1,o1}
\fmf{dbl_curly}{v1,o2}
\fmfv{decor.shape=circle,decor.filled=30,decor.size=.3w}{v1}
\end{fmfgraph*}
\end{gathered}
=
\begin{gathered}
      \begin{fmfgraph*}(80,40)
       \fmfleft{i1,i2}
        \fmfright{o1,o2}
\fmf{photon}{i1,v1,i2}
\fmf{dbl_curly}{o1,v2,o2}
\fmf{dbl_curly}{v1,v2}
\end{fmfgraph*}
\end{gathered}
+
\begin{gathered}
    \begin{fmfgraph*}(80,40)
       \fmfleft{i1,i2}
        \fmfright{o1,o2}
\fmf{photon}{i1,v1}
\fmf{photon}{i2,v2}
\fmf{photon}{v1,v2}
\fmf{dbl_curly}{v1,o1}
\fmf{dbl_curly}{v2,o2}
\end{fmfgraph*}
\end{gathered}
+
\begin{gathered}
      \begin{fmfgraph*}(80,40)
       \fmfleft{i1,i2}
        \fmfright{o1,o2}
\fmf{photon}{i1,v1}
\fmf{photon}{i2,v2}
\fmf{photon}{v1,v2}
\fmf{phantom}{v1,o1}
\fmf{phantom}{v2,o2}
\fmffreeze
\fmf{dbl_curly}{v1,o2}
\fmf{dbl_curly}{v2,o1}
\end{fmfgraph*}
\end{gathered}
+
\begin{gathered}
    \begin{fmfgraph*}(80,40)
       \fmfleft{i1,i2}
        \fmfright{o1,o2}
\fmf{photon}{i1,v1}
\fmf{photon}{i2,v1}
\fmf{dbl_curly}{v1,o1}
\fmf{dbl_curly}{v1,o2}
\end{fmfgraph*}
\end{gathered}
\end{equation*}
\end{fmffile}
  \caption{Gravitional Compton scattering given by the tree-level scattering a massless particle (wavy line) on graviton (curly line) with only gravitational interactions.}
  \label{fig:GCompton}
\end{figure}

In this section we review the gravitational Compton scattering at
tree-level represented in fig~\ref{fig:GCompton} and discussed in
detail in~\cite{Holstein:2006pq,Holstein:2008sx,Bjerrum-Bohr:2013bxa}
and in~\cite{Bjerrum-Bohr:2014lea}. We are interested
in the gravitational Compton scattering of a graviton $g$ from a massless target of spin 0 (scalar $\varphi$), spin $\frac12$ (fermion $\chi$), and spin 1 (photon  $\gamma$). The only interactions that we consider are gravitational interactions.

A remarkable property of the gravitational Compton scattering from a
target $X$ of spin $S$ and mass $M$ (which can be vanishing) is its
factorization onto a product of Abelian QED Compton amplitudes~\cite{Bjerrum-Bohr:2013bxa}
\begin{equation}\label{e:Grav3}
i  \cM_{X^2 G^2} (p_1,k_1,p_2,k_2)={\kappa^2\over 8e^4}\,  {p_1\cdot
k_1\, p_1\cdot k_2\over k_1\cdot k_2}\, A_S^{\rm Compton}(p_1,k_2,p_2,k_1)\, A_0^{\rm Compton}(p_1,k_2,p_2,k_1)\,,
\end{equation}
where $A^{\rm Compton}_S$ is the Compton amplitude associated with
scattering a photon from a target of spin $S$, and $ A^{\rm
  Compton}_0$  is the Compton amplitude obtained by scattering
a photon on scalar target.

We express gravitational Compton amplitudes in the
helicity formalism, using the notation for the polarization dependence
of the external states
\begin{equation}
\cM^{(h_1h_2|\lambda_1\lambda_2)}_{\gamma^2G^2}:=  \cM_\gamma^{\rm G-Compton}(p_1^{h_1}, k_1^{\lambda_1},
p_2^{h_2} ,k_2^{\lambda_2})\,.
\end{equation}
The amplitudes $\cM^{(++|++)}_{\gamma^2G^2}$,
$\cM^{(++|+-)}_{\gamma^2G^2}$, $M^{(++|-+)}_{\gamma^2G^2}$, $\cM^{(+-|++)}_{\gamma^2G^2}$, $\cM^{(-+|++)}_{\gamma^2G^2}$ and
their complex conjugate vanish since the four-gluon tree-level
amplitude
$A^{\rm Compton}_1(p_1^{h_1},k_1^{\lambda_1},p_2^{h_2},k_2^{\lambda_2})$
is zero for these configurations of polarizations as they are not
 MHV amplitudes.  In addition, the
gravitational amplitude $\cM^{(++|--)}_{\gamma^2G^2}$ (and its complex conjugate) vanishes because the scalar amplitudes $A^{\rm Compton}_0(p_1, k_1^{-},p_2,k_2^{-})$ (and its complex conjugate) vanishes for massless scalars.  Thus, the only non-vanishing gravitational Compton amplitudes for photons are, see for example~\cite{Mangano:1990by,Dixon:1996wi,Dixon:2013uaa}.
\begin{eqnarray}\label{e:ggGGhel}
  \cM^{(+-|+-)}_{\gamma^2G^2}&=&{\kappa^2\over8}{\sq[p_1,k_1]^2\an[p_2,k_2]^2\spab[k_2,p_1,k_1]^2\over
    (p_1\cdot p_2)(p_1\cdot k_1)(p_1\cdot k_2)}\,,\\
\cM^{(-+|+-)}_{\gamma^2G^2}&=&
{\kappa^2\over8}{\sq[p_2,k_1]^2\an[p_1,k_2]^2\spab[k_2,p_2,k_1]^2\over
  (p_1\cdot p_2)(p_1\cdot k_1)(p_1\cdot k_2)}\,,
\end{eqnarray}
and their complex conjugates.

For scalar target Compton scattering, the helicity amplitudes derived in~\cite{Bjerrum-Bohr:2013bxa} are given by

%%%%
%
%
\begin{equation}
  \label{e:4pointYM}
A_{0}^{\rm Compton}(p_1,p_2,k_2^+,k_1^+)=-{M^2 \sq[k_1,k_2]^2\over
 k_1\cdot k_2 \, 2k_1\cdot p_1},\qquad
A_{0}^{\rm Compton} (p_1,p_2,k_2^-,k_1^+)={\spab[k_2,p_1,k_1]^2\over
  k_1\cdot k_2\, 2k_1\cdot p_1}\,,
\end{equation}
with the complex conjugated expressions
$A_{0}^{\rm Compton} (p_1,p_2,k_2^-,k_1^-)=(A_{0}^{\rm Compton}
(p_1,p_2,k_2^+,k_1^+))^*$ and
$A_{0}^{\rm Compton} (p_1,p_2,k_2^+,k_1^-)=(A_{0}^{\rm Compton} (p_1,p_2,k_2^-,k_1^+))^*$.

The gravitational Compton amplitude in Eq.~(\ref{e:Grav3}) then reads
in the helicity formalism
\begin{eqnarray}
  \label{e:4pointGrav}
 \cM_{\Phi^2G^2}(p_1,k_1^+,p_2,k_2^+) &=&{\kappa^2\over 16}\,
 {1\over (k_1\cdot k_2) }\,
 {M^4 \sq[k_1,k_2]^4\over (k_1\cdot p_1)(k_1\cdot p_2)} \,, \cr
\cM_{\Phi^2G^2}(p_3,k_1^-,p_4,k_2^+)&=&{\kappa^2\over16}\,{1\over (k_1\cdot k_2)}\,
{ \spab[k_1,p_3,k_2]^2\spab[k_1,p_4,k_2]^2\over (k_1\cdot p_3)(k_1\cdot p_4)}\,.
\end{eqnarray}
Note that for the same reason as in the photon case, the
$\cM_{\Phi^2G^2}(p_1,k_1^+,p_2,k_2^+)$ amplitude vanishes for a massless target, and
in the same way we find the Compton amplitude for massless fermions to be
\begin{equation}
\cM_{\chi^2 G^2}(p_1^+,p_2^-,k_1^+,k_2^-)=\frac{\kappa^2}{16}
\frac{1}{(k_1\cdot k_2)} \frac{\spab[k_2,p_1,k_1]^3 \sq[k_1,p_2]
  \an[p_1,k_2]}{(k_1\cdot p_1) (k_1\cdot p_2)}\,.
\end{equation}

\subsection{The one-graviton tree-level amplitudes}\label{sec:treegrav}

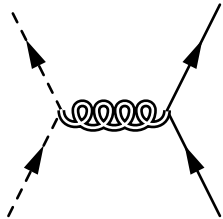
\begin{figure}[ht]
  \centering
\begin{fmffile}{figonegrav}
      \begin{fmfgraph*}(100,80)
       \fmfleft{i1,i2}
        \fmfright{o1,o2}
\fmf{dashes_arrow}{i1,v1,i2}
\fmf{fermion}{o1,v2,o2}
\fmf{dbl_curly}{v1,v2}
\end{fmfgraph*}
\end{fmffile}
  \caption{One graviton exchange between a  massless fields
    and a massive scalar.}
  \label{fig:OneGrav}
\end{figure}

We give the full tree-level one-graviton exchange amplitude  between the massive
scalar $\Phi$ of mass $M$  and the massless
scalar $\varphi$, the photon $\gamma$ and the massless fermion $\chi$.
The massless particles have momenta $p_1$ and $p_2$, the massive
scalar has momenta $p_3$ and $p_4$, with $p_1+p_2+p_3+p_4=0$. The
kinematic invariants are $t=(p_1+p_2)^2$, $s=(p_1+p_4)^2$ and $u=(p_1+p_3)^2$.
\begin{itemize}
\item[$\bullet$] The tree-level gravitational interaction between a  massless scalar
$\varphi$  and a massive scalar $\Phi$ is given by
\begin{equation}
  \label{e:Mscalarm1m2}
  \cM^{(1)}_{\varphi^2\Phi^2}
=-{\kappa^2\over 4} {(s-M^2)(u-M^2)\over t} \,.
\end{equation}
\item[$\bullet$] The tree-level gravitational interaction between a  photon
$\gamma$  and a massive scalar $\Phi$ is given by
\begin{equation}\label{e:treeggsshelicity}
  \cM^{(1)\,(++)}_{\gamma^2\Phi^2}=0;\qquad
  \cM^{(1)\,(+-)}_{\gamma^2\Phi^2}=\kappa^2\,{
    \spab[p_2,p_3,p_1]^2\over 4(p_1+p_2)^2 }= \kappa^2 {(M^4-su)^2\over 4t \spab[p_1,p_3,p_2]^2}\,,
\end{equation}
where the superscript denotes the helicity of the external photons, and we used that $\spab[p_2,p_3,p_1]^2\spab[p_1,p_3,p_2]^2=\big(\tr_-(p_2p_3p_1p_3)\big)^2=(M^4-su)^2$
with equivalent expressions for their complex conjugate helicity
configurations.
The vanishing of the  amplitude when the helicity of the
configuration of the incoming and outgoing photons are the same is  expected from general relativity since two
parallel beams of light do not interact gravitationally~\cite{Tolman}.

\item[$\bullet$]   The tree-level gravitational interaction between a  massless
spin-${1\over 2}$ field $\chi$ and a massive scalar $\Phi$ is given by
\begin{equation}  \cM^{(1)\, (+-)}_{\chi^2 \Phi^2}=\kappa^2{
    (s-u)\over8 t}  \spab[p_2,p_3,p_1]= \kappa^2 \,
  {(s-u)(M^2-su)\over 8 t \spab[p_1,p_3,p_2]}\,.
\end{equation}
with an equivalent expression for the complex conjugate helicity configuration.

\end{itemize}

%--------------------------------------------------------------------------
\section{Integrals}\label{sec:integral}
In this appendix we use the same convention as in the main text
$q=p_1+p_2=-p_3-p_4$, $t=(p_1+p_2)^2$, $s=(p_1+p_4)^2$ and $u=(p_1+p_3)^2$.

\noindent {\bf The infrared divergent integral}

The boxes are defined and evaluated~\cite{qcdloop} as
\begin{equation}\label{e:scalarbox}
  I_4(t,s)= {1\over i\pi^{2-\epsilon} r_\Gamma}\int {d^{4-2\epsilon}\ell \, \mu^{2\epsilon}\over \ell^2
    (\ell+q)^2 (\ell+p_1)^2 ((\ell-p_4)^2-M^2)}\,,
\end{equation}
with the following $\epsilon$ expansion
\begin{multline}
  I_4(t,s)
  =-{1\over t(M^2-s)}\,\left(\mu^2\over
   M^2\right)^\epsilon\,\Bigg[{2\over\epsilon^2}-{1\over\epsilon}(2\log{M^2-s\over
   M^2}+\log{-t\over M^2})\cr
+2\log{M^2-s\over M^2}\log{-t\over M^2}-{\pi^2\over2}+O(\epsilon)\Bigg]\,,
\end{multline}
where
$r_\Gamma=(\Gamma(1-\epsilon))^2\Gamma(1+\epsilon)/\Gamma(1-2\epsilon)$.
The box scalar integral $I_4(t,u)$ is obtained by replacing $s$ by $u$
in the previous expressions.

The massless triangle integral is defined by
\begin{eqnarray}\label{e:masslesstriangle}
  I_3(t)&=&   {1\over i\pi^{2-\epsilon} r_\Gamma} \int {d^{4-2\epsilon}\ell \, \mu^{2\epsilon}\over \ell^2
    (\ell-p_1)^2 (\ell+p_2)^2}
=-{1\over \epsilon^2\, t}\,\left(-t\over \mu^2\right)^{-1-\epsilon}\cr
&=& -{1\over t\,\epsilon^2}- {\log(-t/\mu^2)\over t
  \epsilon}-{(\log(-t/\mu^2))^2\over 2t}+O(\epsilon) \,.
\end{eqnarray}

\noindent{\bf The ultraviolet divergent integral}

The massless bubble integral is defined as
\begin{equation}\label{e:masslessbubble}
  I_2(t)= {1\over i\pi^{2-\epsilon} r_\Gamma} \int {d^{4-2\epsilon}\ell \, \mu^{2\epsilon}\over \ell^2
    (\ell+q)^2} ={1\over\epsilon}+2-\log(-t/\mu^2)+O(\epsilon)\,.
\end{equation}

\noindent{\bf The finite integral}

The massive triangle is given by
\begin{eqnarray}\label{e:massivetriangle}
 I_3(t,M^2)&=&  {1\over i\pi^{2-\epsilon} r_\Gamma}  \int {d^{4-2\epsilon}\ell \, \mu^{2\epsilon}\over (\ell+p_4)^2
    (\ell-p_3)^2 (\ell^2-M^2)}\cr
&=&{1\over t\,\beta}\,\Bigg[4\zeta(2)+2\textrm{Li}_2\left(\beta-1\over\beta+1\right)
+\frac12\,\log^2\left(\beta-1\over\beta+1\right)\Bigg]\,,
\end{eqnarray}
where $\textrm{Li}_2(x)=\sum_{n\geq1} {x^n\over n^2}$ and $\beta^2=1-{4M^2\over t}$.
The non-relativistic limit leads to
\begin{equation}\label{e:trimassNR}
  I_3(t,M^2)\simeq -{1\over 32\pi^2M^2}\, \left(\log\left(-t\over M^2\right)+
    {\pi^2M  \over\sqrt{-t}}\right)\,.
\end{equation}
%
%-------------------------------------------------------------------------

\end{document}